\documentclass[11pt,superscriptaddress]{revtex4-2}

\usepackage{graphicx}
\usepackage{dcolumn}
\usepackage{bm}
\usepackage[table]{xcolor}
\usepackage{graphicx}
\usepackage{adjustbox}
\usepackage{placeins}
\usepackage[T1]{fontenc}
\usepackage{lipsum}
\usepackage{csquotes}
\usepackage{colortbl}
\usepackage{hyperref}
\usepackage{bm}
\usepackage{multirow}
\usepackage{booktabs}
\usepackage{epsfig}
\usepackage{amsmath,amssymb}

\begin{document}

\title{Hierarchical generative modeling for the design of multi-component systems}

\author{Rhyan Barrett}
\thanks{These authors contributed equally.}
\affiliation{Leipzig University, Wilhelm Ostwald Institute for Physical and Theoretical Chemistry, Linnéstraße 2, 04103 Leipzig, Germany}

\author{Robin Curth}
\thanks{These authors contributed equally.}
\affiliation{Leipzig University, Wilhelm Ostwald Institute for Physical and Theoretical Chemistry, Linnéstraße 2, 04103 Leipzig, Germany}
\affiliation{Center for Scalable Data Analytics and Artificial Intelligence (ScaDS.AI), Dresden/Leipzig, Germany}

\author{Julia Westermayr}
\email{julia.westermayr@uni-leipzig.de}
\affiliation{Leipzig University, Wilhelm Ostwald Institute for Physical and Theoretical Chemistry, Linnéstraße 2, 04103 Leipzig, Germany}
\affiliation{Center for Scalable Data Analytics and Artificial Intelligence (ScaDS.AI), Dresden/Leipzig, Germany}

\date{\today}

\begin{abstract}
The functionality of catalysts, enzymes, and supramolecular assemblies emerges not from individual molecules alone, but from the subtle interplay between multiple components arranged in complex systems. Designing such systems is a grand challenge, the combinatorial explosion of possible chemical compositions and spatial arrangements makes brute-force exploration infeasible, while many current generative approaches remain limited to isolated molecules. In this work, we introduce a hierarchical generative optimization framework that overcomes this barrier by coupling a genetic algorithm for configurational search with a generative model for molecular design. This closed-loop approach enables simultaneous refinement of geometry and composition, efficiently steering discovery toward systems with targeted functionality. As a proof of concept, we design catalytic environments for the Claisen rearrangement of p-tolyl ether by optimizing surrounding components around a fixed reference transition-state geometry. Despite this constraint during the search phase, post-hoc validation via Climbing-Image Nudged Elastic Band calculations confirm a 30\% reduction in activation barrier. Beyond this example, our framework provides a general strategy for data-driven discovery of functional multi-component systems, opening the door to automated design of catalysts, enzyme active sites, and advanced materials.
\end{abstract}

\maketitle

\section{Introduction}


A central challenge in molecular discovery is the efficient exploration of chemical space to identify molecules and molecular systems with desired properties \cite{polishchuk2013estimation, reymond2025chemical, von2020exploring}. In cheminformatics, computational methods have become indispensable for this task by enabling virtual screening, property prediction, and data driven prioritization of candidate structures before experimental testing \cite{oliveira2023virtual, gorgulla2023recent, graff2021accelerating}. However, the number of synthetically and functionally accessible compounds remains enormous, and structure property relationships are often highly non-linear, making exhaustive evaluation of predefined libraries increasingly impractical as molecular complexity grows \cite{yang2019analyzing, vamathevan2019applications, reymond2025chemical}. As a result, many current workflows still rely on screening based approaches that rank existing candidates rather than proposing fundamentally new ones \cite{gorgulla2023recent, graff2021accelerating}. This limitation has motivated growing interest in inverse design strategies, particularly generative models that learn molecular representations from data and propose previously unseen structures tailored to target properties \cite{gomez2018automatic, loeffler2024reinvent, gao2020synthesizability}.

Generative machine learning models offer a powerful route to inverse design by learning statistical patterns in molecular data and proposing novel structures beyond those observed during training \cite{sanchez2018inverse, segler2018generating}. They can be trained on simple low-dimensional descriptors such as SMILES strings \cite{jin2018junction, brown2019guacamol} or on three-dimensional molecular structures. Prominent examples include autoregressive models such as G-SchNet \cite{gebauer2019symmetry, gebauer2022inverse} and diffusion-based approaches \cite{corso2022diffdock, jing2022torsional}, which can generate molecular structures directly. When combined with supervised models that capture relationships between chemical structure and target properties, generative models can be steered toward desired regions of chemical space or conditioned to propose molecules with improved properties \cite{gebauer2022inverse, westermayr2023high, therrien2025using, zhavoronkov2019deep}.

Existing case studies, e.g., on the design of molecules using generative models for drug discovery and property optimization \cite{popova2018deep,blaschke2020reinvent,therrien2025using}, demonstrate promising progress, but they remain largely restricted to isolated molecular systems. However, molecules rarely exist in isolation; instead, they function within complex chemical and biological environments. Examples include protein ligand binding sites in enzymes, receptor ligand interactions central to drug discovery, and solvation or molecular complex formation in condensed phases \cite{wei2024structure,cramer2008universal}. In each of these cases, the functionality of the system emerges not only from the properties of the individual components but also from their intricate interactions \cite{pakdel2024high}. Such systems are inherently multi-component, and because the design space of each constituent is already vast, the combined search space grows explosively when considered as a whole \cite{korn2023navigating, liu2023performance}.

Attempting to generate entire multi-component systems directly within a single generative framework \cite{zhang2024efficient, wen2025generative} proves extremely challenging, even when using conditional generative approaches. When designing catalysts or reactive sites, not only the chemical composition needs to be optimized, but also the geometric features like relative orientation, distances, or symmetries of the individual components, to improve a reaction \cite{kiss2013computational}. This combined compositional and configurational space is high-dimensional and heavily constrained by chemical validity and stability requirements. Consequently, rather than generating complete local environments in a single step, recent strategies decompose the problem into smaller, more tractable sub-tasks. Classical hierarchical or multiscale workflows can reduce the problem by first selecting a fixed set of candidate components, then optimizing their spatial arrangement using heuristic search strategies such as Monte Carlo sampling \cite{poole2025accelerating}, evolutionary algorithms\cite{chandraghatgi2024streamlining}, or fragment-based assembly \cite{murray2009rise}. While effective in narrowing down large candidate spaces, these methods remain constrained by the diversity of the initial library and cannot autonomously propose new chemical components. As a result, the search is limited to recombining what is already known, leaving much of chemical space unexplored.

In this work, we combine the best of both worlds in a single framework. In particular, we combine the possibility to go beyond known chemical structures with generative models \cite{elijovsius2025zero} and the multi-component design using hierarchical models and propose a combined hierarchical generative optimization framework that breaks the problem of designing multi-component systems into manageable sub-problems, as illustrated in Figure \ref{fig:method}. In the first stage, a global optimization layer explores the spatial arrangement of components by treating positions, distances, and orientations as tunable parameters. A genetic algorithm searches this configurational space to identify arrangements that minimize a chosen target property. 
Subsequently, a generative modeling layer focuses on the chemical composition of the subunits themselves. By learning from the best-performing environments found in previous iterations, the generative model proposes new molecules with similar or improved stabilizing features, enriching the pool of candidates for subsequent optimization. Together, these two stages form a closed-loop pipeline that iteratively refines both the chemical composition of the subunits and the global structure of the whole system. A key application of our framework is the automated design of local catalytic environments, where subtle changes in electrostatics, dispersion, or hydrogen bonding can dramatically alter reaction rates. As a proof of concept, we apply the method to the Claisen rearrangement reaction of p-tolyl ether \cite{irani2009joint} and reduce the activation barrier by over 30\%, revealing clear interaction motifs that stabilize the transition state, providing further chemical insights into design principles.

\section{Results}
In this work, we apply the hierarchical multi-component optimization framework to the design of local chemical environments for reaction optimization by using the p-tolyl ether Claisen rearrangement reaction as an example. 
The proposed workflow is illustrated in Figure 1 and will be explained in the following.
\subsection{Multi-Component System Optimization Workflow}

\begin{figure}
\centering
\includegraphics[width=\textwidth]{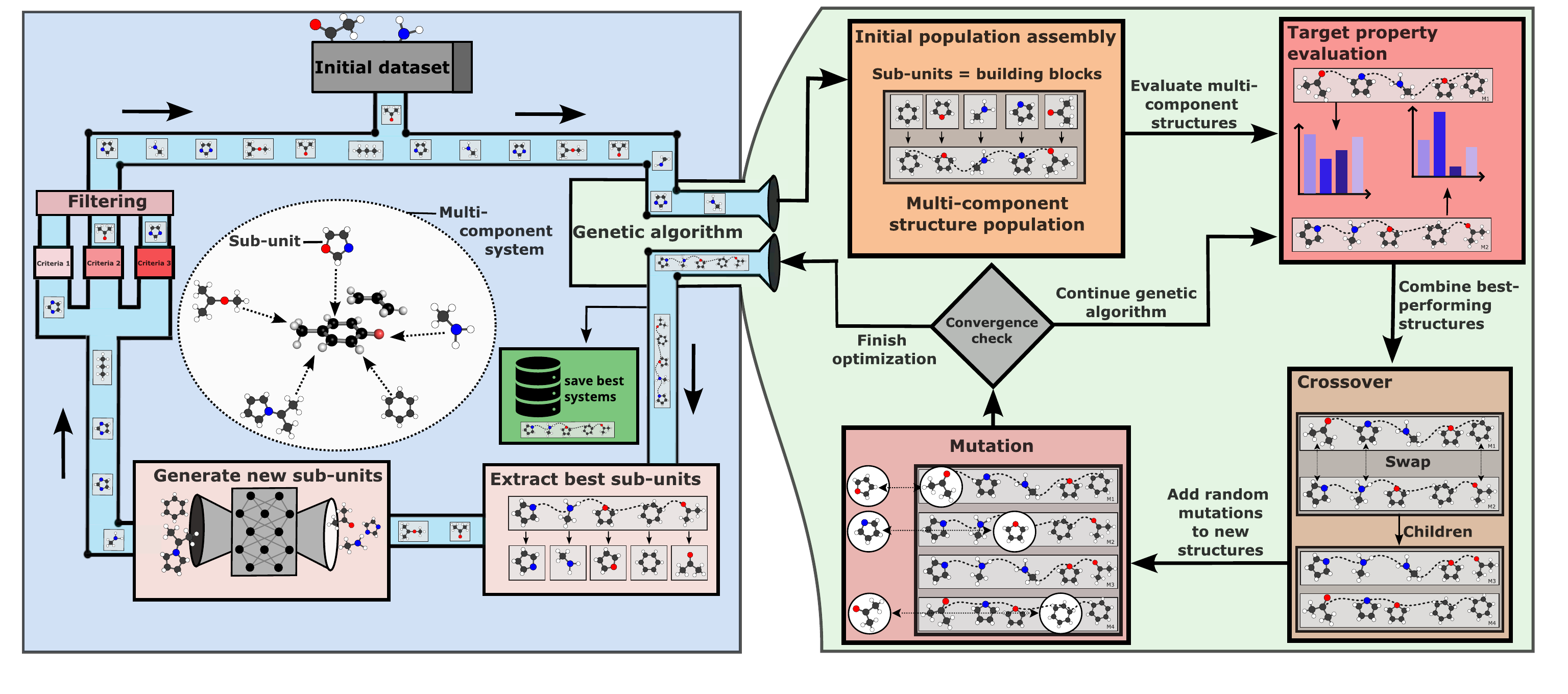}
\caption{Overview of the developed multi-component structure design workflow. Starting with an initial dataset of molecules, a population of environments is constructed. The parameters $x_i$, $\phi_i$ and $\theta_i$, are optimized via a genetic algorithm (right side). Suitable molecules are collected and used as training data for a generative model that generates a new set of possible molecules to form the environment for each position. The loop is repeated until convergence of optimization criterion.}
\label{fig:method}
\end{figure}

The developed workflow (Figure \ref{fig:method}) for designing a multi-component system can be understood as an assembly of several molecular or structural units placed in proximity to one another, as shown in the middle of the left side of Figure \ref{fig:method}. Each configuration is defined by two key elements: (i) a set of candidate building blocks that can occupy the system, and (ii) a global geometry that specifies how these blocks are arranged. The geometry is described by parameters such as distances, orientations, and positions, which together determine how the units are spatially organized, see also Figure \ref{fig:environment_genetic_algo}a. These two elements give rise to two coupled optimization problems: a chemical optimization problem, concerned with the identity and structural motifs of the building blocks, and a geometrical optimization problem, concerned with their spatial arrangement. 

The geometrical optimization problem is addressed using a global optimizer, which we chose to be a genetic algorithm (GA) in this work (illustrated on the right of Figure \ref{fig:method}). At this stage, the pool of candidate molecular units is fixed, and the GA operates only on the parameters defining the overall structure. The GA requires three ingredients: a pool of molecular units, a predictive model for the target property, and the geometrical parameters that describe the arrangement. The optimization proceeds in cycles, illustrated in Figure \ref{fig:method} (right side), referred to as generations, each of which follows four main steps: construction, selection, recombination, and mutation. In the construction step, which is shown in the orange box of Figure \ref{fig:method} an initial population of structures is generated by randomly assigning molecular units to geometrical parameters, and each structure is scored, as illustrated in the red box on the right side of Figure \ref{fig:method}. Scoring is done in this work using a predictive machine learning model trained on quantum chemical properties of organic molecules (see \ref{sec:mace-details} section for details), but in principle, any scoring function can be applied that gives access to target properties.

In practice, as the optimization becomes more exploitative in later generations, the search can enter regions that are poorly represented in the training distribution of the underlying predictive model, leading to chemically implausible proposals. To prevent these from biasing the optimization, all candidates are subjected to chemical validity checks and structural relaxation prior to scoring, see the methods section for details. As foundational models improve in coverage and physical fidelity, we expect such failure modes to become less frequent, increasing the stability of coupled generative optimization workflows.

During selection, which is shown in the box on the bottom right of Figure \ref{fig:method} the best-performing structures with respect to the pre-defined target property are retained, while recombination mixes their parameters to generate new candidates. Mutation (shown in the light-red box of Figure \ref{fig:method}) then introduces small random variations to maintain diversity and avoid premature convergence. Repeating this cycle over many generations allows the GA to refine the global arrangement, propagating favorable configurations while still exploring new possibilities. The full details of the GA can be found in the methods section and Figure \ref{fig:environment_genetic_algo}.

After the geometrical optimization with the GA converges, the molecular units that most frequently appear in high-performing systems are collected and used to retrain a generative model that was previously trained on an initial pool of potential candidates (see left part of Figure \ref{fig:method}). By biasing towards promising systems, this model proposes new candidate units with features similar to those found in the previous best-performing structures. A new round of geometrical optimization is performed with the new pool of candidates. By alternating between genetic optimization of geometry and generative modeling of building blocks, the workflow systematically improves both the arrangement and composition of the system until the target property reaches convergence. In principle, one could replace this two-stage loop with a fully generative end-to-end approach that directly proposes complete multi-component assemblies. However, maintaining chemical validity and stability of such large assemblies after construction is substantially more challenging than for small molecules; we therefore use the genetic algorithm as a search layer and the generative model to produce smaller, more stable molecules.

\subsection{Case study: Application to Local Catalytic Environments}

\begin{figure}
\centering
\includegraphics[width=0.6\textwidth]{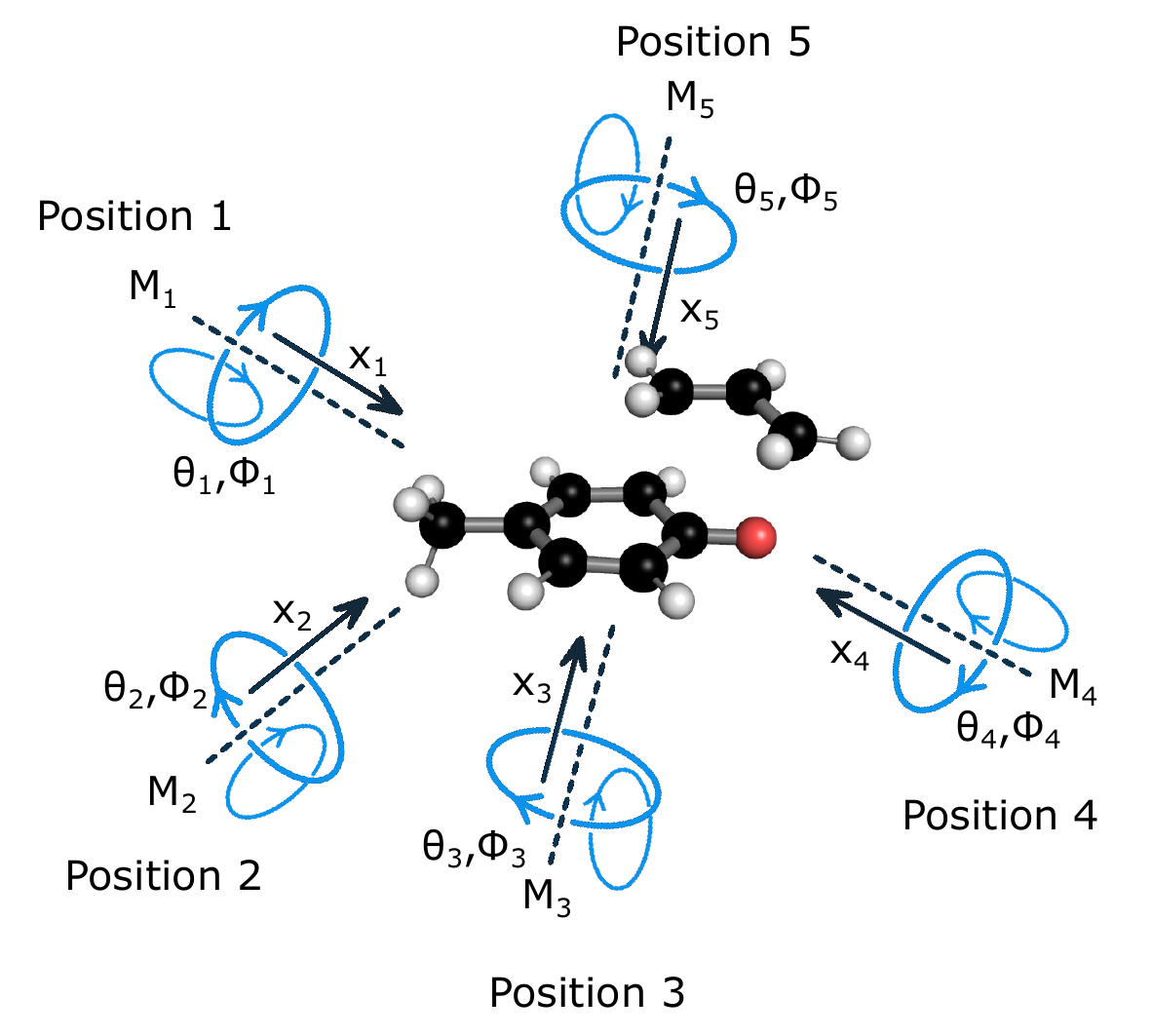}
\caption{Transition-state test system and environment parameterization. The transition-state geometry is shown together with five predefined placement directions (positions $j=1,\dots,5$) for the surrounding molecular subunits. One subunit $M_j$ is assigned to each position $j$ and is parameterized by $(x_j,\theta_j,\phi_j)$, corresponding to its distance from the transition state and two orientation angles that specify its placement and orientation.}
\label{fig:example_structure}
\end{figure}

Building on the general workflow described above, we applied the approach to the design of a local catalytic environment for the Claisen rearrangement of p-tolyl ether. The environment was represented by five molecular subunits placed along predefined vectors surrounding the transition state, as can be seen in Fig.~\ref{fig:example_structure}. At initialization, one subunit is assigned to each of the five positions by random sampling from the SPICE dataset, which provides the candidate pool used throughout the workflow. Initially for each vector $j$, the subunit distance $x_j$ is drawn uniformly from a feasible range and its orientation is randomized via the angles $(\theta_j,\phi_j)$; the GA then optimizes both molecular identity and these placement parameters. We chose five subunits as a simple binding pocket like setup that allows multiple simultaneous interactions with the reacting complex while keeping the number of subunits around the transition state feasible; the number of subunits can be changed by adding or removing placement vectors. Each vector was parameterized by one distance and two angular degrees of freedom, defining how a molecule can be positioned and oriented relative to the reaction center. A pool of candidate molecules was assigned to each vector, from which the GA constructed and optimized local environments.

Although we consider five positions around a single transition state here (Fig.~\ref{fig:example_structure}), the framework is general. For other applications, the number and placement of positions are chosen to match the underlying scaffold structure, while the pool of molecular subunits is selected to reflect the domain of interest. 
As an example using a protein binding site, positions can be defined by pocket anchor points, and candidate subunits can be amino-acid variants at these anchors, with scoring based on predicted binding affinity together with selectivity constraints against off-targets.

In this example we define the target property to be the interaction energy between the vacuum transition state and the environment, defined as
\begin{equation}
\delta E = E_{TS_{env}} - (E_{env} + E_{TS}),
\end{equation}
where $E_{TS_{env}}$ is the energy of the full system, and $E_{env}$ and $E_{TS}$ correspond to the isolated fragments, i.e., the environment and the transition state of the target reaction, respectively. Assuming the transition-state geometry within the new environment remains close to the original one, negative interaction energies indicate stabilization of the transition state and thus a reduction in activation barrier. We note, however, that $\delta E$ is evaluated for a fixed reference (gas-phase) transition-state geometry and is used as an ML screening objective; it is therefore not identical to the true activation barrier obtained from NEB/DFT, where the reactant and product minima are optimized and the barrier is determined along the resulting minimum-energy pathway, so the environment can stabilize the reactant and the barrier region to different extents. The final improvement in stabilization energy is later confirmed using quantum-chemical calculations and minimum energy pathway optimization using Nudged Elastic Band (NEB) calculations. Therefore, we use the message-passing atomic cluster expansion model (MACE), in particular the MACE-OFF23 model \cite{kovacsMACEOFFTransferableShort2025b} that is trained of millions of organic molecules. This model is finetuned on reference calculations of randomly generated environments computed at the wB97M-D3BJ level of theory \cite{mardirossian2016omegab97m, grimme2011effect}, where the environment subunits are sampled from the SPICE dataset \cite{eastmanSPICEDatasetDruglike2023} as a source of candidate fragments; the fine-tuning labels are newly computed DFT interaction energies for the constructed Claisen transition-state environment complexes to refine the predictions for our specific system. While we use the fine-tuned model in this work, the optimization loop can in principle also be run with a non fine-tuned foundational model. Further details regarding the reference calculations and the finetuning can be found in the supporting information (SI) in section S1.

After each optimization cycle, the subunits most frequently present in high-performing environments are collected to guide the generative stage. For this, SiMGen~\cite{elijovsius2025zero} is used. SiMGen is a generative extension of MACE~\cite{kovacs2025mace}, which proposes a new set of candidate molecules conditioned on top-performing structures. 
Using these new candidates, the genetic optimization is repeated. This alternating procedure of geometric optimization and generative expansion is performed until convergence is reached and no further improvement in $\delta E$ is observed.

The most promising environments from the final iteration are then validated against quantum-chemical calculations and analyzed to provide chemical insights and potentially guide further design and optimization processes. 

\begin{figure}[htbp]
\centering
\includegraphics[width=\textwidth]{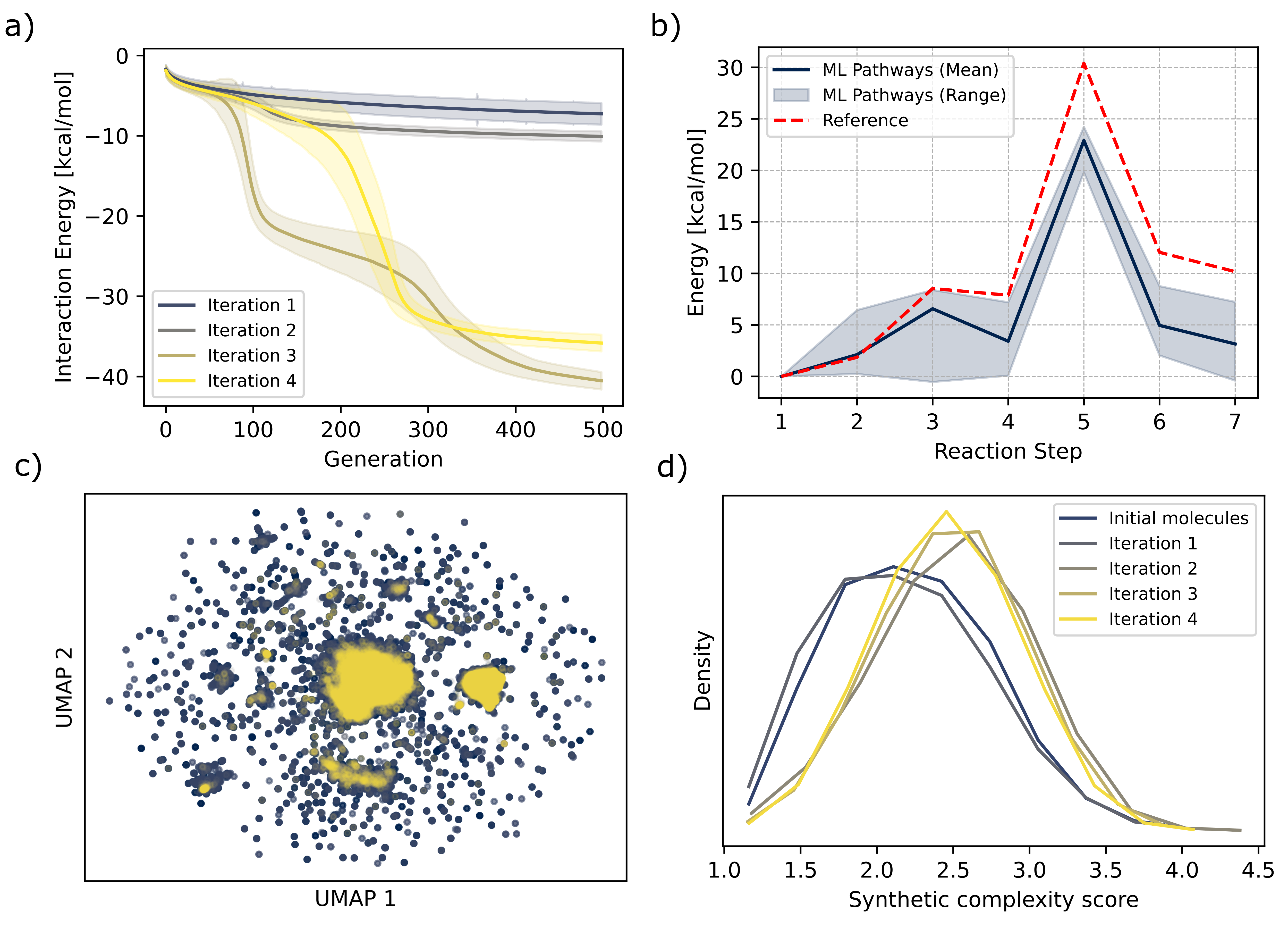}
\caption{ a) Progress of the genetic algorithms. Solid lines represent the mean interaction energy of all generated local chemical environments with respect to the reference transition state and shaded areas show the standard deviation. b) Reaction energy profile of the non-catalyzed reaction in vacuum (red) and of the reaction in 40 designed environments using NEB calculations. The blue line depicts the mean value obtained from all local environments, and the shaded area the standard deviation. c) UMAP-projection of SOAP descriptors of the environment molecules, colored by iteration (the same legend as in c) applies). d) Synthetic complexity score distributions of the environment molecules colored by generative iteration.}
\label{fig:results}
\end{figure}

The results can be seen in Figure \ref{fig:results}. In total, four optimization runs were conducted until the interaction energy did not improve further. Here, four optimization runs refers to four outer-loop iterations of the hierarchical workflow, where each iteration consists of a GA run of 500 generations followed by a generative update of the subunit pool. We stopped the outer loop once the mean predicted interaction energy no longer improved compared to the previous iteration (Fig.~\ref{fig:results}a); an additional iteration did not yield further gains in the ML score. Panel a) shows the mean predicted interaction energy as a function of generations of the GA for each run. As can be seen, each individual run of the GA shows a downward trend in the target property and convergence in the late stages of the GA. While iteration 1 converges towards a value of approximately -7.4 kcal/mol, iteration 2 reaches a value of approximately -10.2 kcal/mol. 
This improvement can be attributed to the generative biasing step, which shifts the distribution of candidate molecules toward those more potent at stabilizing the transition state. 
Iteration 3 converges at a much lower average interaction energy of about -40.6 kcal/mol, and the development of the interaction energy shows a stepwise decrease. Analysis of the molecules at each position shows that the stepwise decrease in interaction energy is caused by the algorithm discovering a highly optimal configuration, which results in extensive exploitation. Iteration 4 converges at around -35.7 kcal/mol, which is a bit higher than that of iteration 3. Further runs did not show improvement, and hence the process was stopped after this iteration.
To validate whether the modified local environments lead to an increased stabilization of the transition state, hence increased reaction rate, we validate the reaction using the NEB method \cite{henkelman2000climbing} at the Density Functional Theory (DFT) level to confirm the reduction in activation energy on a set of the best performing environments from the last iteration. Here, the reported approximate activation energy is defined as the energy difference between the reactant minimum and the highest-energy climbing image along the CI-NEB path, with the environment kept fixed, providing a consistent comparative barrier estimate across environments. The NEB method optimizes a reaction path and searches for the minimum energy pathway for each local environment. We have selected 40 top-performing local chemical environments for validation. The resulting energy curves from the minimum energy paths of the different designed local chemical environments can be seen in Figure \ref{fig:results}b). As can be seen, the activation energy is reduced from ~30 kcal/mol to ~20 kcal / mol on average when comparing the original (vacuum) pathway with those of the top 40 environments. These calculations validate the use of the finetuned MACE version to assess the propensity of local chemical environments to stabilize the transition state.

To investigate how and why certain molecules are more likely to lead to an energy stabilization, we analyze the change in the distribution of environment molecules during the different iterations. Therefore, each position in the local environment is treated separately and molecules are represented using the Smooth Overlap of Atomic Positions (SOAP) \cite {bartokRepresentingChemicalEnvironments2013b} descriptor. SOAP embeds a molecular structure in a high-dimensional space, providing a way to mathematically represent the chemical structure space spanned by a set of molecules. A Uniform Manifold Approximation and Projection (UMAP) \cite{mcinnes2018umap} dimensionality reduction was performed for visualization with details specified in the SI. section. UMAP takes the structural SOAP descriptor and transforms it to a lower-dimensional space, while preserving as much of the global structure of the data distribution as possible. Figure \ref{fig:results} c shows the distribution of the molecules in this new structure space spanned by the two UMAP components. It can be seen that the molecules become less diverse with increasing iteration and span a narrower chemical space. This finding is reasonable as certain molecular motifs are more likely to stabilize the transition state than others and are identified during the iterative cycles. Figure S4 in the SI shows the chemical space spanned for each position in the local chemical environment separately. To assess whether molecules are still valid and synthesizable, we additionally plot the distribution of the synthetic complexity scores (SCScore) \cite{coleySCScoreSyntheticComplexity2018} that provides a measure of how reasonable a molecule is, which is illustrated in Fig. \ref{fig:results}d. A higher SCScore relates to more complex synthesis, smaller values are generally preferable. As can be seen, after the first iteration, the distribution is almost identical to the original data set but shifts slightly to larger values in iteration 2, where it remains nearly constant in further iterations. Notably, some of the highest performing candidates as shown in Fig. \ref{fig:chem_analysis}d likely fall into the upper tail of the SCScore distribution, indicating a trade-off between predicted catalytic stabilization and synthetic accessibility. This suggests that while the model can identify strongly stabilizing environments, a subset of these designs may be challenging to realize experimentally. In future works, this can be addressed by complementing SCScore with retrosynthesis predictions (e.g., predicted route length or feasibility) and incorporating these as additional filters or as a secondary objective during generative biasing.

This shift in predicted synthetic complexity is likely caused by the transition from known literature-based molecules towards machine learning-generated molecules, which may compromise the synthesizability as a side result. Still, given the general much lower synthesizability and likeliness of molecular structures generated with generative machine learning in recent studies, the SCScore can be considered fairly low \cite{gao2020synthesizability,westermayr2023high}. This result is further supported by distribution of bond lengths and angles plotted in Figure S3 in the SI comparing generated and original molecular structures.

\subsubsection{Chemical insights and design principles}

\begin{figure}[htbp]
\centering
\includegraphics[width=\textwidth]{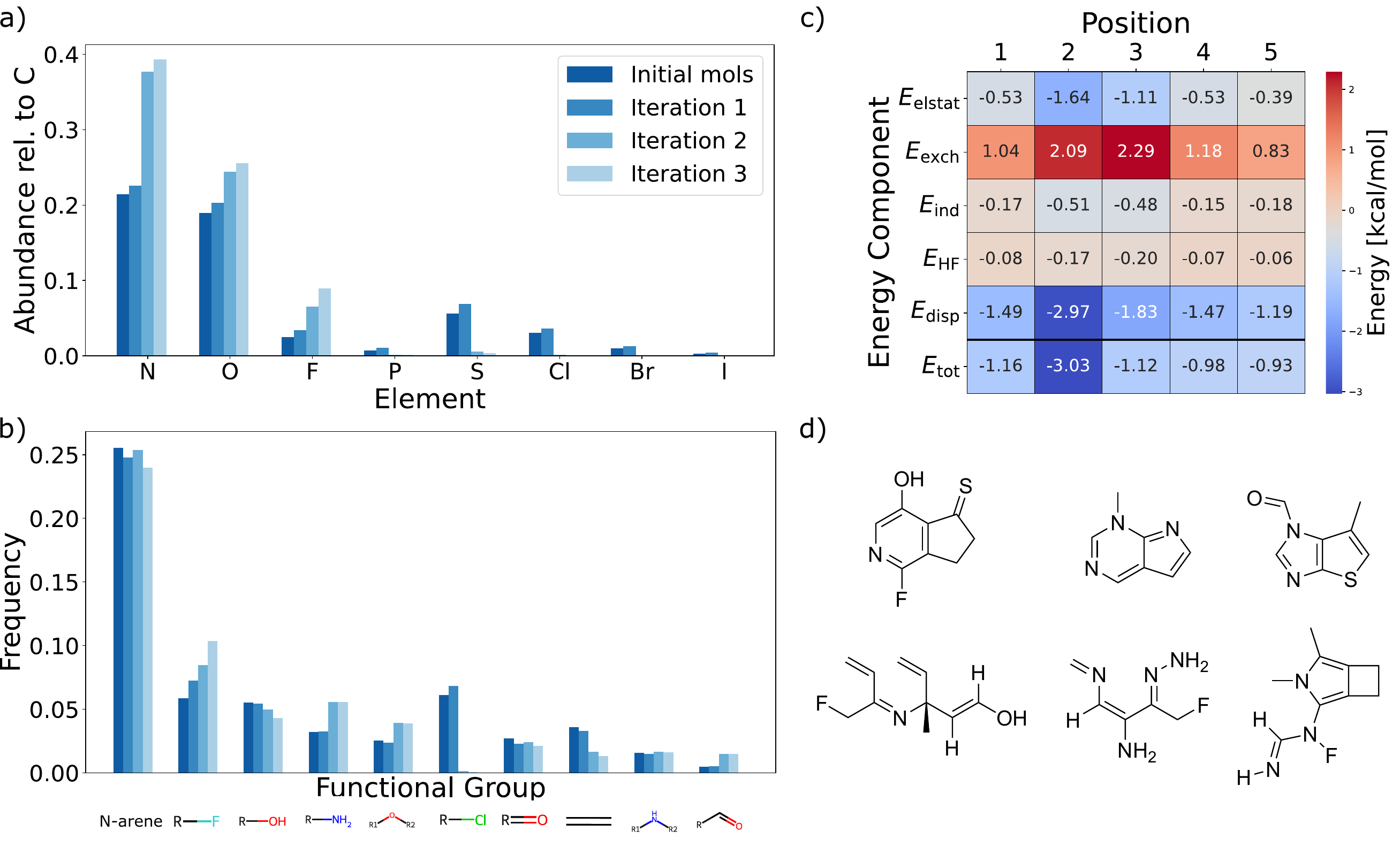}
\caption{Development of a) the atom type distribution and b) the functional group distribution of the sets of environment molecules during the generative biasing steps. c) Average SAPT interaction energy components between the environment molecules of the 40 environments chosen for further analysis and the transition state for each of the five environment vectors. d) Selection of environment molecules from the last iteration.}
\label{fig:chem_analysis}
\end{figure}

Finally, to gain a better understanding of what influences the reaction rate, chemical analysis of the molecules that lead to a stabilized transition state is performed. Therefore, after each generative phase converged, molecules that lead to the highest stabilization energy increase are collected. As these molecules are also used to bias the generative model, the elemental composition of the environment molecules changed significantly with each iteration and generation step, which is shown in Fig.~\ref{fig:chem_analysis}a). 
As can be seen, the proportion of the elements fluorine, nitrogen and oxygen relative to carbon increased in the course of the iterations. In contrast, the presence of larger halogens (Cl, Br, I) and third-row heteroatoms (P, S) declined during the iterations. No explicit constraints or penalties were applied to these elements; instead, this shift emerges from the iterative selection-and-biasing loop, where the GA preferentially retains environments with favorable predicted interaction energies and the generative stage is then conditioned on the subunits enriched among these top-performing environments, thereby reinforcing the compositional trends across iterations. The increase in the former set of atoms (F, N, O), being the most electronegative atoms, likely points to a larger electrostatic interaction between the transition state and the environment molecules. 
To further monitor changes in the molecules that make up the local chemical environment of the reaction, different functional groups were investigated. These were selected based on structural motifs that were observed via visible inspection of molecules that led to improved stabilization energy. 
The development of the frequency of the ten most common functional groups found in the datasets is shown in Fig.~\ref{fig:chem_analysis}b. The most common functional groups are N-heterocyclic aromatic groups, followed by fluorinated groups, alcohols, and primary amines. As already discussed, the amount of fluorinated groups increases, and the amount of chlorine drastically decreases. Primary amine groups, ether groups, and aldehydes become more frequent, while carbon-carbon double bonds become less frequent. However, apart from the enrichment in fluorine and the decline in chlorine content, no specific functional group stands out as the one that is best-suited for catalyzing the reaction.

The environment is represented by vectors that constrain the positioning of molecules around the transition state and exhibit distinct behaviors during the genetic optimization cycles and the generative biasing stages. The relative orientation of the different vectors to the transition state might lead to different types of interactions, e.g. hydrogen bonding interactions occur if molecules are positioned close to the oxygen atom of the transition state. 
To gain insights into the nature of non-covalent interactions between the environment molecules and the transition state, and how they differ between the five different vectors that constrain the system, we performed Symmetry Adapted Perturbation Theory (SAPT) \cite{jeziorskiPerturbationTheoryApproach1994} calculations of the best-performing environment structures of the last iteration. SAPT decomposes the non-covalent interaction energy between two fragments into electrostatics, induction, dispersion, and exchange contributions.

The calculations were done in a pairwise fashion, with the transition state and one of the environment molecules as fragments at the PBE0/aug-cc-pvdz \cite{adamo1999toward} level of theory.
The resulting average interaction energy contributions for each position are shown in Fig.~\ref{fig:chem_analysis}c. The SAPT calculations show that the strongest average interactions occur between molecules at position~2 (Fig.~\ref{fig:example_structure}), with total interaction energies about three times larger compared to the other positions.
This strong attractive interaction at position~2 is driven by large dispersion interactions and electrostatic interactions, as well as a large induction component. Position~2 is located close to the aromatic ring of the transition state. Based on the energy profile of the interaction and the orientation of position~2, it is expected that strong interactions stem primarily from $\pi$-stacking interactions. The large amount of N-heterocyclic aromatic molecules in the datasets supports this hypothesis. Panel c in Figure 3 further shows that electrostatic interactions are strongest at positions 2 and 3, which can be explained by interactions with the electron-rich oxygen atom of the transition state, for example, via hydrogen bonding. The other environment vectors are positioned further away from the oxygen atoms (see Figure 4a), leading to stronger electrostatic interactions. Figure \ref{fig:chem_analysis}d displays selected molecules from the final iteration that are capable of engaging in pi-stacking interactions (top row) and hydrogen bonding interactions (bottom row), both of which have been shown to favorably stabilize the transition state.
The use of hydrogen-bond donors as catalysts for the Claisen rearrangement reaction has already been reported in the literature \cite{uyedaEnantioselectiveClaisenRearrangements2008a}.  

\section{Discussion}

In this work, we develop a hierarchical generative framework for efficient, property-driven design of multi-component molecular systems. The approach decomposes the design task into two coupled but tractable subproblems: global spatial optimization via a genetic algorithm and local subunit generation via a molecular generative model. By alternating between these stages, the workflow can move beyond a fixed initial library while optimizing the spatial context in which molecules act. In the present proof-of-principle, the transition-state geometry is kept fixed during optimization, such that the framework identifies environments that stabilize a given reference structure and thereby isolates the effect of the surrounding multi-component arrangement on reaction energetics. Extending the approach to cases where the environment actively reshapes the reaction pathway will require coupling the search to iterative transition-state or reaction-path optimization, which represents a natural direction for future work.

To enable property-driven optimization, the workflow requires a scoring function, which we implement here as a predictive machine learning model. This dependence requires reliable supervised models, since inaccurate predictions can bias the search toward misleading candidates. This risk can be mitigated by employing ensemble models or explicit uncertainty quantification. We have shown that molecules generated in this work are still valid and synthesizable. 
Additional screening and biasing using predictive models for synthesizability or scoring functions therefore can further improve the generative stage. Penalization of high predictor uncertainty can help ensure that generated candidates are both realistic and practical.

Compared to conventional approaches, which often rely on fragment-based assembly or heuristic search without generative exploration, our method combines the strengths of both optimization and generation. While many applications of generative modeling have so far been restricted to small molecules, this framework extends the concept to complex environments such as catalytic pockets, enzyme active sites, and supramolecular assemblies, where interactions emerge from both chemical composition and spatial arrangement.

Future work will aim to combine the approach with more robust machine learning potentials, improved neural network architectures, or explicit uncertainty modeling to enhance predictive accuracy. Recently developed foundational models \cite{woodUMAFamilyUniversal2025, levineOpenMolecules20252025} provide avenues for improved predictions. In addition, expanding the framework to other use cases, such as the design of functional enzyme pockets or tailored supramolecular assemblies, will help test the applicability of the workflow in broader contexts.

\section{Methods}

\subsection{Genetic Algorithm}
In the following, we describe how the GA sets up and optimizes the arrangement of molecules forming the local chemical environment of a transition state. The GA determines the type of molecule and its orientation/position.

\subsubsection{Environment Construction and Parameterization}
\begin{figure}[htbp]
\centering
\includegraphics[width=\textwidth]{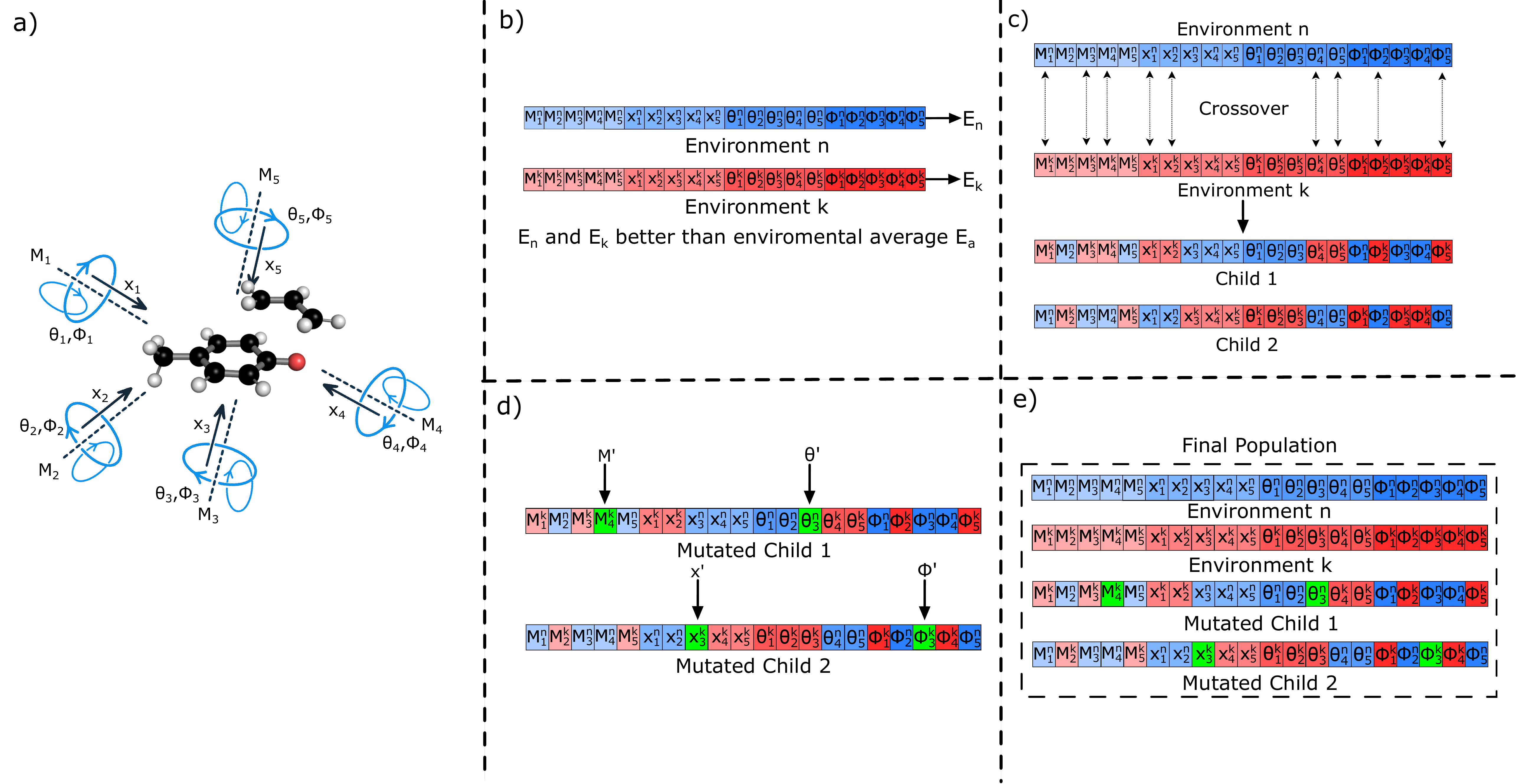}
\caption{a) Transition state of the Claisen rearrangement of p-tolyl ether with surrounding vectors characterized by four parameters: molecule index, intermolecular distance, and two angular coordinates describing the orientation of the molecules making up the local chemical environment. b) Each environment, defined by its parameters, is evaluated through an energy prediction; environments performing above average advance to the next stage of the genetic algorithm. c) New environments are generated from parent environments by randomly selecting and exchanging parameters. d) With a certain probability, random parameter mutations are introduced into each environment. e) A new population is formed from the modified environments, continuing the evolutionary cycle.}
\label{fig:environment_genetic_algo}
\end{figure}

The first step in the genetic optimization phase is to construct a population of local chemical environments around a given reaction under investigation. In this work, we chose the environment to be five molecules that could be attached to, e.g., enzyme active sites, and position them along a set of predefined vectors, denoted by $S$, around a central reaction molecule (see Figure \ref{fig:environment_genetic_algo}a). This design is flexible; additional vectors can be added or removed to adjust the number of surrounding molecules as needed.

For each vector, $j$, a candidate molecule is selected from the SPICE dataset \cite{eastmanSPICEDatasetDruglike2023} (see Figure S1 in the SI for details on the data set), which was also used and re-computed to finetune the MACE-OFF23 model and train the initial generative model, and assigned to a vector. The position along the vector is denoted by \( x_j \) and is typically chosen from a uniform distribution over a feasible range. Additionally, each molecule can be arbitrarily rotated. The rotation is parameterized by two angles, \(\theta_j\) and \(\phi_j\), which describe the orientation of the molecule within the rotation group \( SO(3) \). Including the molecular identifier \( i \), the complete set of parameters for a molecule assigned to vector \( j \) is given by:

\begin{equation}
\left\{ (i, x_j, \theta_j, \phi_j) \;:\; j \in S \right\}.
\end{equation}

In our implementation, the use of five vectors yields five independent sets of parameters for each candidate environment. The placement of the five vectors around the transition state of the investigated reaction is shown in Figure \ref{fig:environment_genetic_algo}a.

\subsubsection{Initial Population and Evaluation}
An initial population of candidate environments is generated by repeating the above process \( n \) times. For each environment, a molecule is randomly selected for each vector \( j \), and its corresponding parameters \((x_j, \theta_j, \phi_j)\) are generated randomly within their respective domains. This stochastic initialization ensures diversity in the initial search space.

The performance of each environment is then evaluated using property predictions with the finetuned MACE-OFF23 model in this work. The key metric is the interaction energy between the transition state and the environment, which is defined in equation (1).

\subsubsection{Overlap Phase (Crossover)}
Following evaluation, the GA proceeds by selecting the top-performing half of the population based on the defined score. These selected individuals form the basis for the crossover (or overlap) phase, which is designed to combine favorable traits from different environments (see Figure \ref{fig:environment_genetic_algo}c). 

For each vector \( j \), consider two parent environments with parameter sets:
\begin{equation}
\{(i^1, x_j^1, \theta_j^1, \phi_j^1), \quad (i^2, x_j^2, \theta_j^2, \phi_j^2)\}.
\end{equation}
A child environment is produced by randomly shuffling the parameter sets from the two parents. Specifically, for each parameter in the child’s configuration \( C_j = \{(i^C, x_j^C, \theta_j^C, \phi_j^C)\} \), the value is chosen from the corresponding parameter of either parent with equal probability:

\begin{equation}
P(i^C = i^1) = P(i^C = i^2) = 0.5,
\end{equation}

and similarly for \( x_j \), \(\theta_j\), and \(\phi_j\). This procedure is applied independently for every vector \( j \), ensuring that the resulting child environment inherits a mix of characteristics from both parents. In the present work, this crossover operator was kept fixed and was not systematically varied. We chose this parameter-wise recombination because it provides a simple and unbiased mixing of parental configurations while remaining straightforward to implement. Although alternative crossover schemes could also be used, exploring them was beyond the scope of the present study. The crossover process is iterated until the population size is restored to its original number, thereby maintaining diversity while concentrating beneficial traits.

\subsubsection{Mutation Phase}
To avoid premature convergence to local optima, a mutation step is incorporated into the GA (see Figure \ref{fig:environment_genetic_algo}d). During this phase, each parameter of the child environment is subjected to mutation with a fixed probability \(\gamma\). The mutation introduces random changes, thereby allowing exploration of new regions in the parameter space.

Mathematically, for any given parameter:

\begin{equation}
P\bigl((i^C, x_j^C, \theta_j^C, \phi_j^C) \rightarrow (i_M^C, x_j^C, \theta_j^C, \phi_j^C)\bigr) = \gamma,
\end{equation}

\begin{equation}
P\bigl((i^C, x_j^C, \theta_j^C, \phi_j^C) \rightarrow (i^C, x_j^C, \theta_j^C, \phi_j^C)\bigr) = 1 - \gamma.
\end{equation}

Here, the mutation could affect the molecular index \( i^C \), the position \( x_j^C \), or the rotation angles \(\theta_j^C\) and \(\phi_j^C\). In this example the molecular index was effected by the mutation. New values for a mutated parameter are drawn from the same distributions used during the initial population generation. The mutation rate \(\gamma\) is chosen to balance the need for exploration (diversity) with the preservation of beneficial configurations. This mutation process is applied to every molecule in every candidate environment, forming the next generation of the population. The genetic optimization cycle, consisting of evaluation, crossover, and mutation, is repeated for 500 generations per hierarchical iteration. The outer hierarchical loop is then repeated until no further improvement in the best interaction energy is observed across iterations.

\subsection{Generative Modeling}
Once the GA has converged and a set of high-performing environments has been identified, the next phase involves generative modeling, \textit{i.e.}, molecular design. In this phase, the molecular configurations from the best-performing environments are collected and aggregated. Specifically, for each predefined vector \( j \), the top-performing 5,000 molecules (40 \%) were aggregated by traversing the generations of the genetic algorithm starting from the last generation. This number was chosen to ensure a large enough chemical variability for the biasing step of the generative model \cite{westermayr2023high}. 
The data sets, which capture the structural characteristics that contributed to a reduction in the activation energy, are combined to form a comprehensive dataset. 

To generate the new set of candidate structures for each position \(j\), the aggregated molecules from the previous iteration are used as reference data for a generative model. We used SiMGen \cite{elijovsius2025zero}, a diffusion-like generative model based on similarity kernels. SiMGen is a zero-shot local generator that moves atoms along the gradient of a time-dependent similarity score. In our runs, the descriptors $\{\boldsymbol{\chi}_i\}$ are the node representations extracted from the MACE-OFF23 model\cite{kovacs2025mace}. The similarity score is computed using a Gaussian radial basis function and is computed between all node descriptors in the reference $\{\boldsymbol{\chi}_i\}$ and generated structure $\{\boldsymbol{\chi}_j\}$. The kernel is defined as

\begin{equation}
k\!\left(\boldsymbol{\chi}_i,\boldsymbol{\chi}_j;t\right)
= \exp\!\left[
  -\,\frac{\bigl\lVert \boldsymbol{\chi}_i - \boldsymbol{\chi}_j \bigr\rVert^2}
          {2\,\sigma(t)^2}
  \right] \, ,
\end{equation}
where $\sigma(t)$ is the kernel width dependent on time. Initially, this similarity kernel only includes an average restorative force that acts on the heavy atom backbone, guiding generated structures towards the average of the reference distribution. After this stage, hydrogen atoms are added in a separate refinement step and the resulting molecules are briefly relaxed using a machine-learned interatomic potential. All other parameters follow the public SiMGen defaults.
The latent space in this work was provided by a pre-trained MACE-OFF23 model \cite{kovacsMACEOFFTransferableShort2025b}. The use of this kernel-based generative model comes with the advantage that retraining of the generative model after each iteration is not required, only a set of reference structures has to be provided.
By using the top-performing 40 \% of sub-units from the previous iteration to sample new sub-units, the distribution of candidates is biased towards molecules with structural features more favorable to improve the target property.

To maintain a high quality of environment molecules, especially in later iterations, we applied filtering and relaxation steps (see section S2 in the SI for details). Therefore, the generated molecular geometries were optimized using MACE-OFF23 and molecules that did not converge were filtered out. Additional filtering criteria include charge neutrality, valid bond orders, and a minimum inter-heavy atom distance of 1.05 \AA. As the optimization process relies on MACE-OFF23, finetuned on the SPICE data set (see Figure S2 for scatter plots indicating the model's generally good performance) and not additionally finetuned on the designed molecules, a limitation of this approach could be that potential molecules might not converge due to high errors in forces. Still, due to the foundational knowledge underlying the model being trained on millions of organic molecules, this limitation is expected to concern only few molecular structures as a large range of chemical space is already covered. Additional active learning and sampling steps could reduce the risk of missing molecules due to potential holes in the fitted energy surface.

\subsection{MACE Details}
\label{sec:mace-details}

MACE models combine $E(3)$-equivariant message passing with an atomic–cluster–expansion (ACE) basis ~\cite{drautz2019atomic,batatia2022mace, witt2023acepotentials}. The total energy is expressed as a sum of learned atomic contributions; when forces are included, they are computed by taking the derivative of the MACE model with respect to the atomic positions.

For energy predictions, the architectural parameters were taken from the MACE-OFF medium model \cite{kovacs2025mace}. A radial cutoff of $r_c=5~\text{\AA}$ was used to construct the local graph. The ACE includes explicit two and three body terms with angular channels up to $l_{\max}=3$. Distances were expanded using radial basis functions, specifically Bessel functions, angular components used spherical harmonics with coupling between them computed with the Clebsch Gordan coefficients, and atom types were encoded with learned embeddings. In order to finetune the energy prediction model a learning rate of $10^{-3}$ was used along with a learning decay of $5\cdot 10^{-10}$. The loss consists of the root mean squared errors of the energy. The number of epochs was set to 200. A scatter plot showing the performance of MACE finetuned on the SPICE data is shown in Figure S2 in the SI.

\subsection{SAPT(DFT) Details}
SAPT \cite{jeziorskiPerturbationTheoryApproach1994} was employed to decompose the intermolecular interaction energies between the transition state and the molecular sub-units of the top-performing environments. SAPT is a perturbative method in which noncovalent interaction energies between two fragments are partitioned into physically meaningful components, namely electrostatics, exchange, induction and dispersion. \cite{jeziorskiPerturbationTheoryApproach1994}.
This allows for a more detailed interpretation of the physical interactions contributing to the stabilization of the transition state. 
In SAPT(DFT), the fragments are described by Kohn-Sham orbitals.
The SAPT(DFT) calculations were carried out using PSI4 \cite{parrish2017psi4} at the PBE0 \cite{adamo1999toward}/aug-cc-pvdz level of theory. PBE0 was used in contrast to PBE for NEB calculations as only single point calculations were required and PBE0 is considered to be more accurate. For every of the 40 local chemical environments optimized with NEB (see section 4.6), five SAPT(DFT) calculations were carried out. Each calculation involved two fragments: the transition state structure of the Claisen rearrangement and one of the five molecules constituting the environment. Splitting each environment into five pairwise SAPT(DFT) calculations allowed a more detailed decomposition of the interaction energies while reducing the overall computational cost.

\subsection{NEB Details}\label{sec:neb}
NEB calculations were performed using the Atomic Simulation Environment (ASE) \cite{larsen2017atomic} interfaced with ORCA 6.0.0 \cite{neese2025software} and xtb \cite{bannwarth2019gfn2}. Prior to DFT (PBE/TZVP), we used GFN2-xTB for optimization. The NEB calculations were performed using climbing image. Due to the high computational cost of DFT, a convergence threshold of 0.15 eV/A was applied. Geometry optimizations employed the BFGS \cite{dai2002convergence} algorithm with a convergence criterion of 0.15 eV/A. DFT calculations were carried out with the PBE functional \cite{perdew1996generalized} in combination with the def2-TZVP basis set. To reduce the number of geometry optimization and NEB iterations atoms in the surrounding environment were kept fixed. 

\section*{Acknowledgements}
This work is funded in parts by the Deutsche Forschungsgemeinschaft (DFG) -- Project-ID 443871192 - GRK 2721: "Hydrogen Isotopes $^{1,2,3}$H". The authors acknowledge the ZIH TU Dresden, the URZ Leipzig University, and Paderborn Center for Parallel Computing (PC2) for providing the computational resources to conduct this study. We thank Jakob Schramm for quantum chemical reference calculations (NEB) of the allyl-\textit{p}-tolyl ether Claisen rearrangement reaction and help with analysis. In preparing this manuscript, the authors utilized ChatGPT and Perplexity AI in the text to improve the language and overall clarity. All aspects of the manuscript were carried out by the authors, who assume responsibility for the work.

\section*{Author Contributions}
R.B. conceived the original idea. R.B. and J.W. planned and designed the project. R.B. implemented the code for the genetic modeling, and R.C. implemented the generative modeling and ran the generative modeling loops. R.C. and R.B. performed analyses of the outputs. All authors contributed to discussions throughout the project. R.B., R.C., and J.W. wrote and refined the manuscript.

\section*{Competing Interests}
The authors declare no competing interests.

\section{Code Availability}
The code for the generative model can be found at https://github.com/RokasEl/simgen. The associated MACE code and models for finetuning  can be found at https://github.com/ACEsuit/mace.

\section{Data Availability}
All data and code generated for this work can be found at: \url{https://doi.org/10.6084/m9.figshare.30265882}

\section*{Supplementary Information}

\section{Dataset construction and MACE-OFF23 finetuning}

The dataset for finetuning the medium MACE-OFF23 model \cite{kovacs2025mace}, as well as for the first iteration of the GA, was constructed from the SPICE dataset \cite{eastmanSPICEDatasetDruglike2023}. The SPICE dataset provides an extensive set of small organic molecules, covering various chemical compositions and conformations. A filtering process was performed first using the following criteria, molecules were restricted to contain only elements implemented in the MACE-OFF models (H, C, N, O, F, P, S, Cl, Br, and I atoms), an even electron count and a size of 6–14 heavy atoms. Restricting the maximum number of heavy atoms to 14 ensures that during the genetic optimization cycles, the molecular sub-units collide less frequently during the assembly of valid populations.
Distributions of size and atom types can be seen in Fig. \ref{fig:distribution}.
We define heavy atoms as any non-hydrogen atom.
\begin{figure}[htbp]
    \centering
    \includegraphics[width=\textwidth]{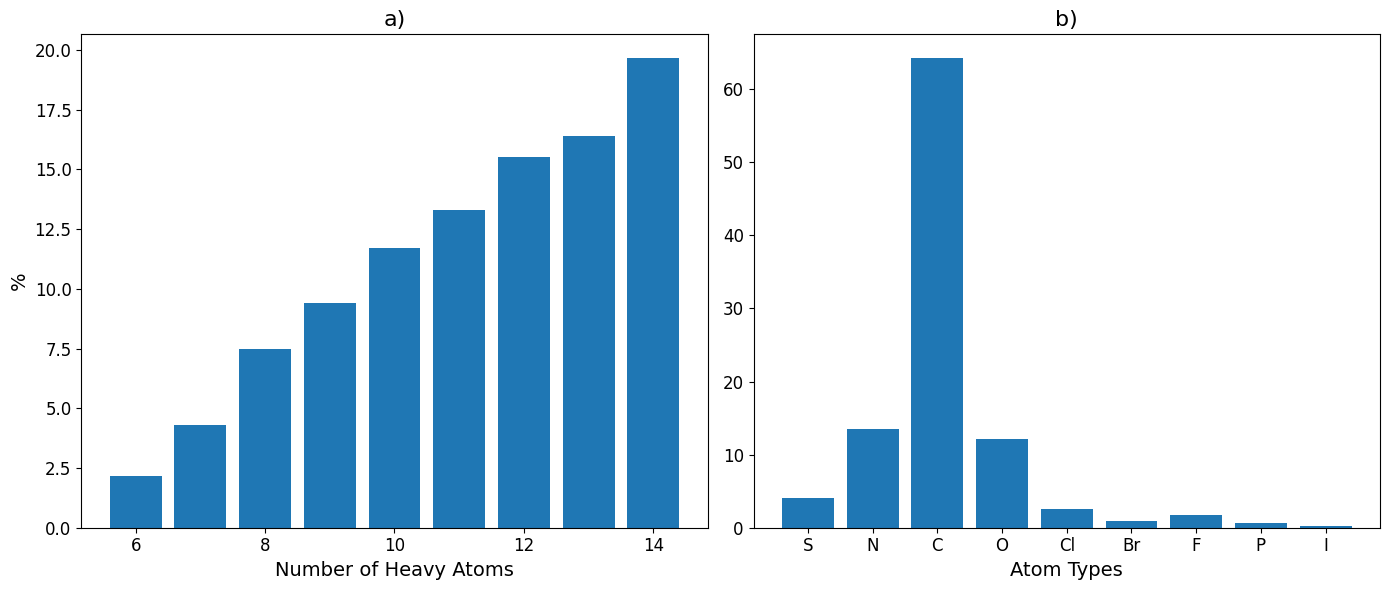}
    \caption{This figure shows the distribution of a) the number of heavy atoms and b) the atom type
    distribution of the initial molecular dataset obtained by sampling from the SPICE dataset.}
    \label{fig:distribution}
\end{figure}

To construct the initial dataset for training the supervised model, randomly selected molecules from selected SPICE molecules were combined with randomly generated orientations and distances. Distances were sampled between 2.5 and 3.5 \AA, while orientations were uniformly sampled from 0 to 360 degrees. 

\section{MACE NEB Study}

\begin{figure}[htbp]
    \centering
    \includegraphics[width=\textwidth]{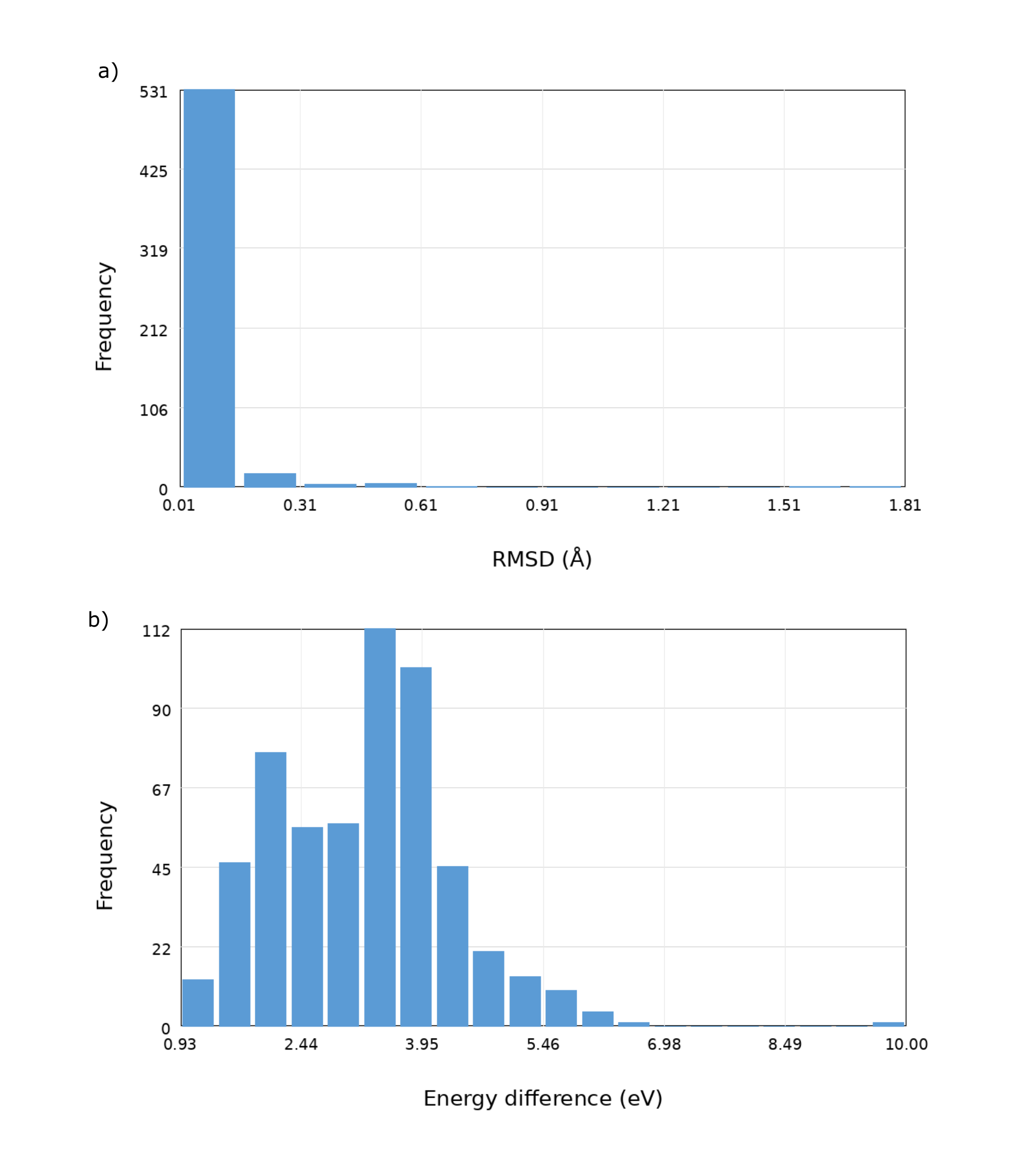}
    \caption{
        (a) Histogram of RMSDs between transition-state region structures obtained from MACE-OFF and ORCA NEB calculations for candidate environments from the final generative cycle. For each system, a corresponding image along the NEB pathway is extracted and the structures are aligned using the Kabsch algorithm prior to RMSD evaluation. 
        (b) Histogram of energy differences between the two methods across all images in the NEB.
        }
    \label{fig:distribution}
\end{figure}

Additionally, we compare transition-state geometries obtained from the MACE-OFF model NEB calculations and from ORCA NEB calculations for candidate environments selected in the final generative cycle. For each system, a corresponding image along the NEB pathway is extracted and the atomic structures are aligned using the Kabsch algorithm to remove overall translation and rotation. The root mean square deviation (RMSD) between atomic positions is then computed as a measure of structural similarity (Fig.~\ref{fig:distribution}a).

The resulting RMSD distribution is strongly peaked at low values, with the vast majority of structures exhibiting RMSDs below $\sim 0.2\,\text{\AA}$ and a pronounced maximum in the first bin ($\approx 0.01$--$0.1\,\text{\AA}$). This indicates that, for most candidates, the MACE-OFF model reproduces the DFT refined ORCA transition-state geometries with high fidelity. Only a small number of outliers extend to larger deviations (up to $\sim 1.5$--$1.8\,\text{\AA}$), corresponding to cases where the ML model and ORCA optimization diverge more significantly. Overall, the distribution demonstrates that MACE-OFF captures the local transition-state geometry accurately across the final candidate set, while highlighting that accurate structural agreement does not necessarily translate directly into quantitative agreement in activation energies.

To further assess this, we compute the differences in activation energies between the two methods for each system, defined as the difference in barrier heights ($\Delta\Delta E$) extracted from the corresponding NEB energy profiles (Fig.~\ref{fig:distribution}b). This provides a direct measure of agreement in the predicted energetics across all images.

Despite this strong structural agreement, the activation energies predicted by MACE-OFF can deviate from those obtained with ORCA. This is likely due to differences in the underlying levels of theory, as well as the fact that the present system may not be well represented in the training distribution of the MACE-OFF model. Fine tuning the model, as discussed in the original work, can help alleviate these discrepancies.

\subsection{Reference calculations}

To reduce the computational cost of dataset generation, interaction energies were computed from pairwise DFT calculations between the reaction transition state and each molecule in the environment. Calculations were performed at the range-separated hybrid functional wB97M-D3BJ \cite{mardirossian2016omegab97m} level, which incorporates Grimme’s D3BJ dispersion correction \cite{grimme2011effect} to account for long-range interactions, in combination with the def2-TZVP basis set, a triple zeta valence basis with polarization functions.

For each pair, the total energy of the combined system (transition state plus a single environment molecule) was calculated, together with the energies of the two individual molecules. The interaction energy was defined as the difference between the combined energy and the sum of the two monomer energies. Finally, the total interaction energy of the environment was obtained by summing all pairwise contributions across the set of environmental molecules.

\subsection{Model finetuning}

\begin{figure}[htbp]
\centering
\includegraphics[width=\textwidth]{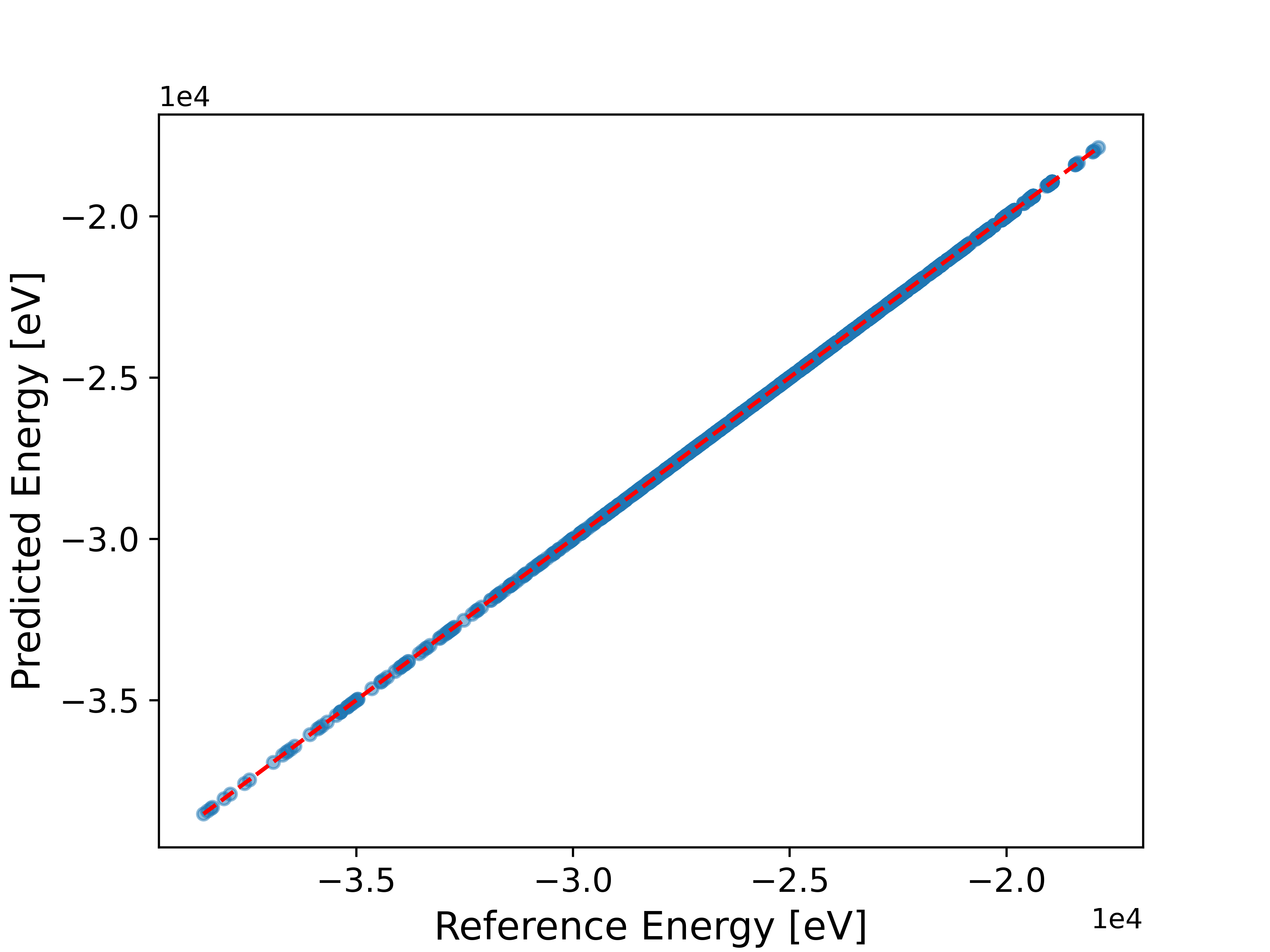}
\caption{Reference energies of the test set versus predicted energies of the finetuned MACE-OFF23 medium model.}
\label{fig:scatter}
\end{figure}

The MACE-OFF23 medium model was used as the starting point for finetuning on the dataset of 36,667 reference calculations using the method defined above. The training was performed for 200 epochs with a learning rate of $10^{-3}$. The final model was evaluated on the test set by predicting the energies and comparing to the reference values from the DFT calculations. The corresponding scatterplot is shown in Fig. \ref{fig:scatter}. 

\section{Post-processing and molecular filtering}

To ensure that the molecules generated by the SiMGen model were both chemically valid and physically meaningful, molecules were filtered based on predefined criteria. As a first step, all candidate structures were subjected to geometry optimization using the MACE-OFF model. Only those molecules that could be optimized were retained. This criterion ensured that molecules correspond to genuine stationary points on the potential energy surface rather than unstable configurations. In addition, each structure was sorted based on the number of electrons, such that only molecules with an even number of electrons were retained, ensuring closed-shell states and excluding radicals or other open-shell species. Structural plausibility was further enforced by having minimum interatomic distance thresholds, preventing the inclusion of molecules with unrealistically short bond lengths or atomic clashes. All remaining structures were required to be convertible into valid molecular graphs using RDKit\cite{landrum2013rdkit} with consistent bond orders and chemically valid connectivity. 

\section{Generative modeling}

\begin{figure}[htbp]
\centering
\includegraphics[width=\textwidth]{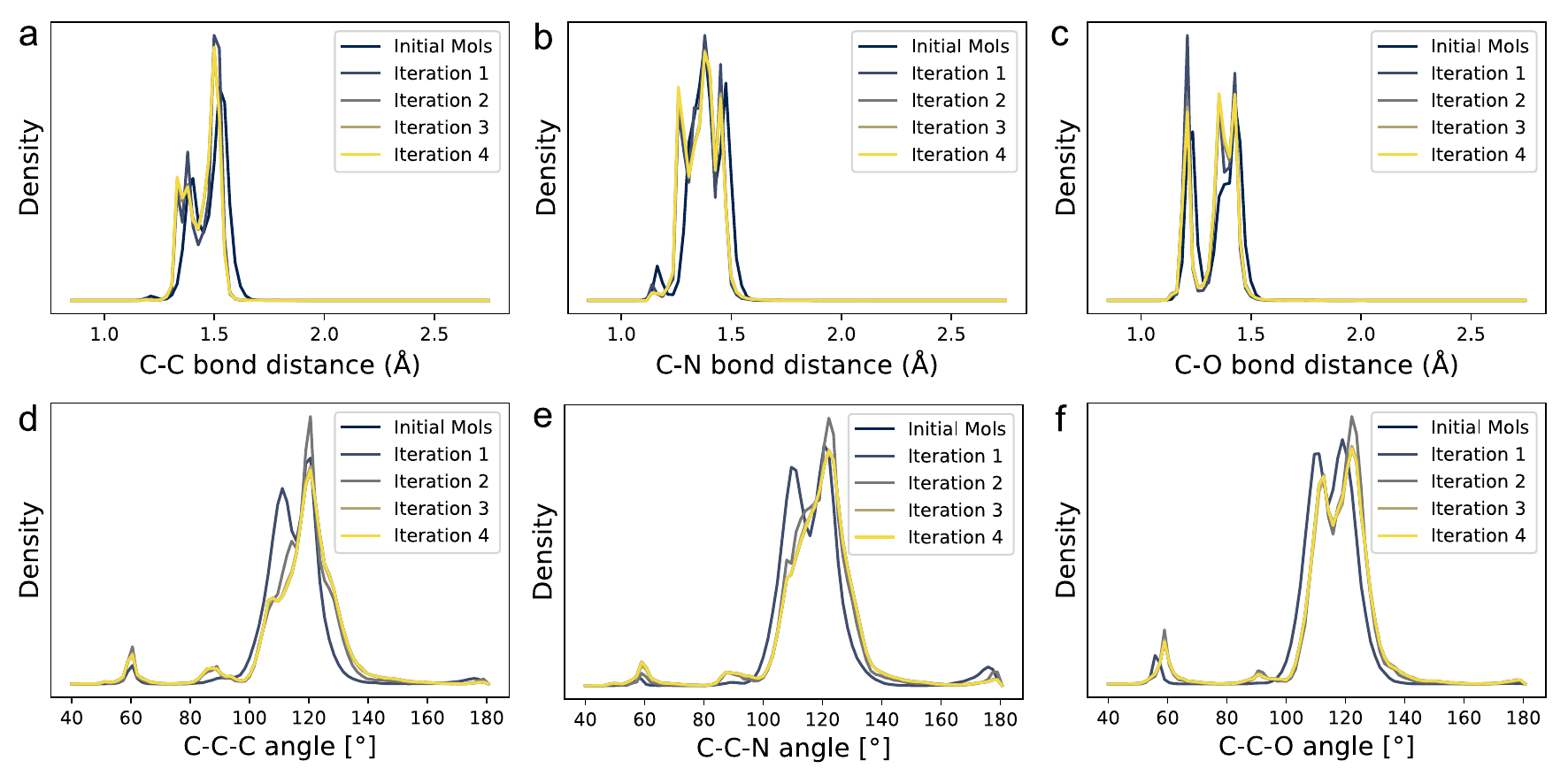}
\caption{Distributions of a) C-C, b) C-N, and c) C-O bonds as well as d) C-C-C, e) C-C-N, and f) C-C-O bond angles.}
\label{fig:bond_length_angles}
\end{figure}

To further validate generated molecules with SiMGen~\cite{elijovsius2025zero}, bond angles and bond lengths were calculated and compared across successive iterations, with the first iteration constructed from the SPICE dataset. We have selected C-C, C-N, and C-O bond lengths as well as corresponding angles as examples to compare distributions. The distributions of the selected bond-lengths and angles can be seen in \ref{fig:bond_length_angles}. In all cases, distributions remained closely aligned with those of the original dataset, indicating that the generated molecules preserve structural characteristics. 

\section{Dimensionality reduction}

\begin{figure}[htbp]
\centering
\includegraphics[width=\textwidth]{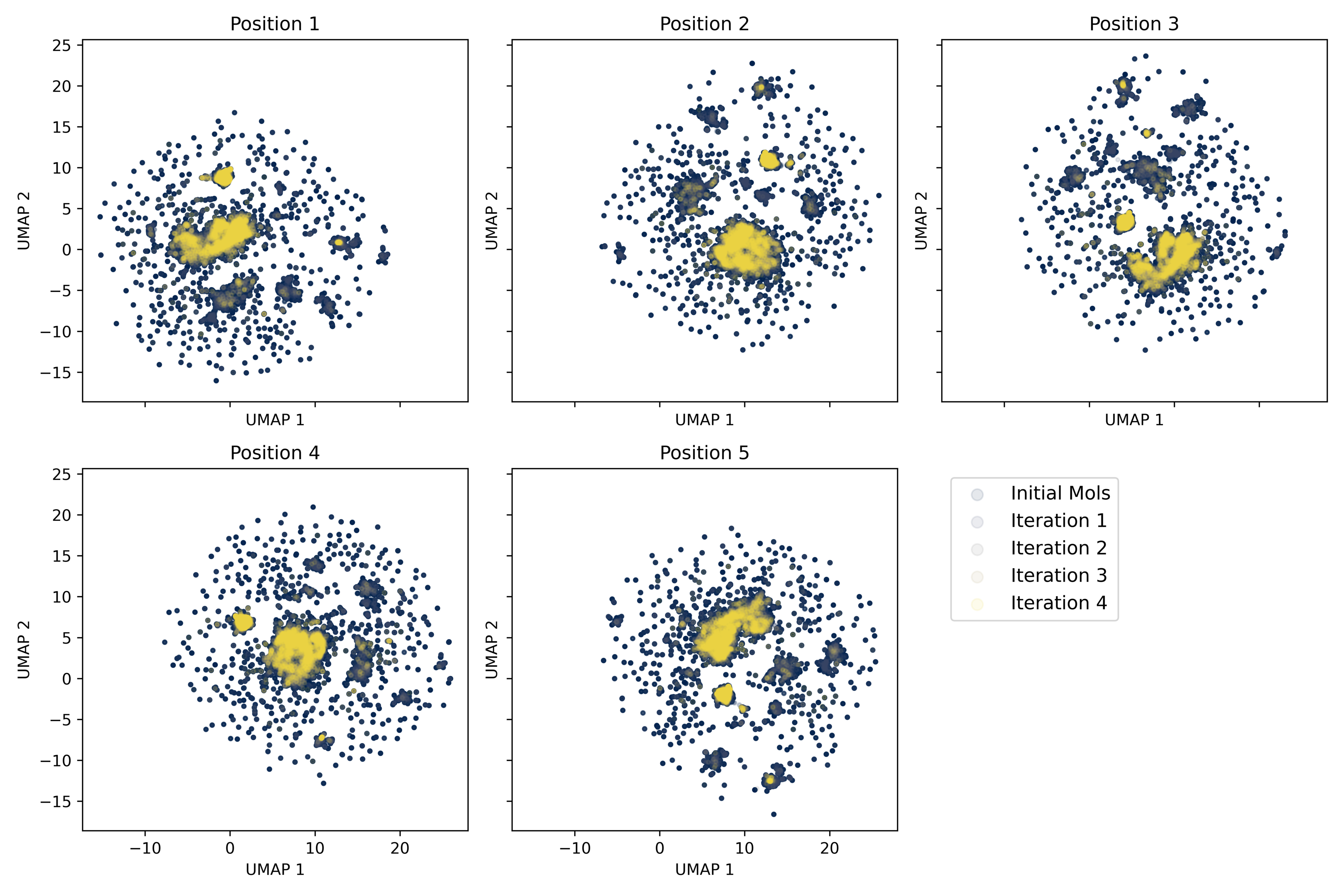}
\caption{UMAP projections of the initial molecule dataset sampled from SPICE and the datasets generated during the different iterations of our workflow for each of the five local positions.}
\label{fig:umap_per_position}
\end{figure}

To analyze how the distribution of generated molecules evolves across chemical space, the Smooth Overlap of Atomic Positions (SOAP) descriptor \cite {bartokRepresentingChemicalEnvironments2013b} was calculated for the initial molecular dataset and for the datasets of extracted sub-unit molecules after each iteration of the workflow. SOAP represents atomic environments by expanding neighbor densities into radial and spherical basis functions and comparing them via smooth overlaps, yielding rotation- and permutation-invariant descriptors for molecular geometries. We used four radial basis functions, a maximum spherical harmonics angular momentum of three and a cutoff radius of three \AA. Due to variable numbers of atoms per molecule, we used inner-product aggregation of the atomic descriptors to obtain consistent descriptor-shapes for each molecule descriptor. For the computation of the soap descriptors, we used the dscribe python package \cite{himanenDScribeLibraryDescriptors2020}.
To visualize the high-dimensional descriptor space in a lower-dimensional space, we performed Uniform Manifold Approximation and Projection (UMAP) \cite{mcinnes2018umap} dimensionality reduction. UMAP works by constructing a weighted graph of local neighborhoods to approximate the manifold structure of high-dimensional data and then optimizing a low-dimensional embedding that preserves those neighborhood relationships. For the UMAP projection of the SOAP descriptors, we used the umap python package \cite{mcinnes2018umap-software}. The number of neighbors was set to 50, the minimum distance parameter was set to 0.1 and we chose to use the cosine similarity as a distance metric. 
For the UMAP projection plot in Figure 2c of the main paper, performed a joint dimensionality reduction of all SOAP descriptors and colored the points by iteration of the workflow. 
The UMAP projection of each individually, colored by workflow iteration is shown in Figure \ref{fig:umap_per_position}.

\section{Claisen Rearrangement Reaction Lewis Structures}
\begin{figure}[htbp]
\centering
\includegraphics[width=0.5\textwidth]{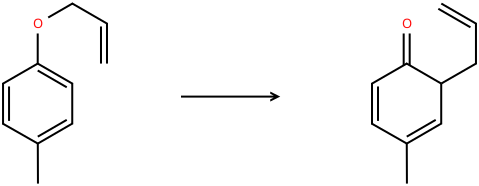}
\caption{Lewis structures of reactant and product structures in the studied reaction}
\label{fig:lewis}
\end{figure}

\section{Choice of SAPT(DFT) funcitonals}
The SAPT(DFT) calculations were carried out using PSI4 \cite{parrish2017psi4}. While the reference interaction-energy calculations used wB97M, this meta-GGA functional is not supported in the PSI4 SAPT(DFT) implementation. PBE0 \cite{adamo1999toward} was therefore used as a supported hybrid functional for SAPT(DFT). The aug-cc-pVDZ basis was selected because augmented correlation-consistent basis sets include diffuse functions, which are important for accurately describing long-range polarization effects and the induction and dispersion contributions in noncovalent interactions."
PBE0 was used in contrast to PBE for NEB calculations as only single point calculations were required and PBE0 is considered to be more accurate. 

\section{Progress of Genetic Algorithm including Fifth Iteration}

\begin{figure}
    \centering
    \includegraphics[width=0.5\linewidth]{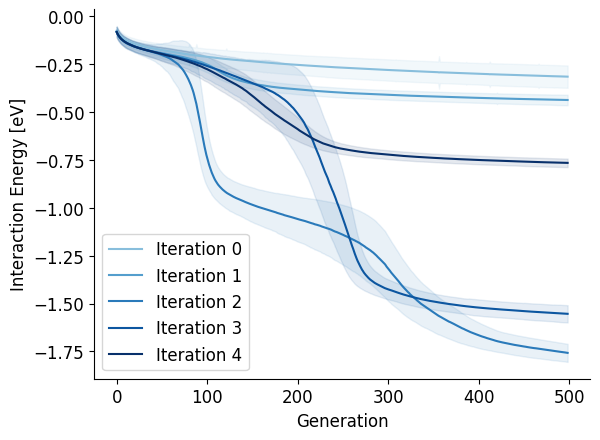}
    \caption{Progress of the genetic algorithms. Solid lines represent the mean interaction energy of all generated local chemical environments with respect to the reference transition state and shaded areas show the standard deviation.}
    \label{fig:including_fifth_iteration}
\end{figure}

To further assess convergence of the outer hierarchical loop, we performed an additional fifth iteration consisting of another full GA optimization over 500 generations followed by generative updating of the subunit pool (Fig.~\ref{fig:including_fifth_iteration}). As shown, the fifth iteration did not improve the mean interaction energy beyond the best values obtained in iterations 3 and 4, but instead converged to a higher final value. We therefore stopped the iterative procedure after four reported iterations. In addition, later iterations showed a growing fraction of unstable or invalid candidate structures, further reducing the effective yield of useful environments.

\bibliography{refs}

\end{document}